\documentclass[prb,aps]{revtex4}% Physical Review Letters

\usepackage{graphicx,color}% Include figure files
\usepackage{dcolumn}% Align table columns on decimal point
\usepackage{bm}% bold math

\newcommand{\bracket}[2]
{\left\langle {#1} \right.\left| {#2} \right\rangle}
\begin{document}

\title{Mesoscopic Stern-Gerlach spin filter by nonuniform spin-orbit interaction}

\author{Jun-ichiro Ohe$^{1}$, Masayuki Yamamoto$^{2}$,
Tomi Ohtsuki$^{1}$ and Junsaku Nitta$^{3}$}

\affiliation{%
$^{1}$Department of Physics, Sophia University,
Kioi-cho 7-1, Chiyoda-ku, Tokyo 102-8554, Japan \\
$^{2}$I. Institut f\"{u}r Theoretische Physik, Universtit\"{a}t Hamburg,
Jungiusstrasse 9, 20355 Hamburg, Germany \\
$^{3}$ NTT Basic Research Laboratories, NTT Corporation, Atsugi-shi,
Kanagawa 243-0198, Japan, and
CREST-JST, 4-1-8 Honcho, Kawaguchi, Saitama 332-0012, Japan}
\date{\today}% It is always \today, today,
             %  but any date may be explicitly specified

\begin{abstract}

A novel spin filtering in two-dimensional electron
system with nonuniform spin-orbit interactions (SOI) is
theoretically studied.
The strength of SOI is modulated perpendicular to the charge current.
A spatial gradient of effective magnetic field due to the nonuniform SOI causes
the Stern-Gerlach type spin separation.
The direction of the polarization is perpendicular to the current
and parallel to the spatial gradient.
Almost 100 \% spin polarization can be realized even without applying
any external magnetic fields
and without attaching ferromagnetic contacts.
The spin polarization persists even in the presence of randomness.

\end{abstract}


\maketitle

Considerable attention has been devoted to the manipulation of
the electron spin
in semiconductor systems.\cite{Wolf,Zutic}
To control the spin of electrons in semiconductors, the spin-orbit
interaction (SOI) due to the lack of the inversion symmetry in 
two dimensional electron system (2DES) is
quite useful since its strength can be controlled by an additional
gate voltage.\cite{Nitta,Engels}
The spin field-effect transistor proposed by Datta and Das
is a hybrid structure of
ferromagnetic electrodes (FM) and a semiconductor (SM)
2DES channel.\cite{Datta}
This hybrid device, however, requires a complex and careful fabrication process.
Furthermore, spin injection efficiency from FM into SM is poor because of 
a conductance mismatch between FM and SM.\cite{Schmidt}
Therefore, it is desired that spin polarized carriers in semiconductor channels
can be generated and manipulated without attaching any ferromagnetic contacts
and without applying any external magnetic fields. 
Several devices based on the SOI are proposed
to have spin polarized carriers in
semiconductor 
channel.\cite{Murakami,Kiselev,Pareek,Koga,Ioniciou,NittaA,Zhou}
From experimental point of view,
the spin polarization mechanism should be robust against disorder. 

One of the historical experiments of the spin separation
is the Stern-Gerlach experiment.\cite{SG}
They have considered a particle with spin propagating
through the nonuniform magnetic field.
Because the derivative of the magnetic field plays a role of the
spin-dependent potential,
particles with up and down spins are accelerated
in opposite directions.
However, for electrons, this effect is hard to
observe because of the effect of the Lorentz force acting
on a electron beam in transverse directions.\cite{Mott}

\begin{figure}[htb]
\begin{center}
\includegraphics[scale=0.5]{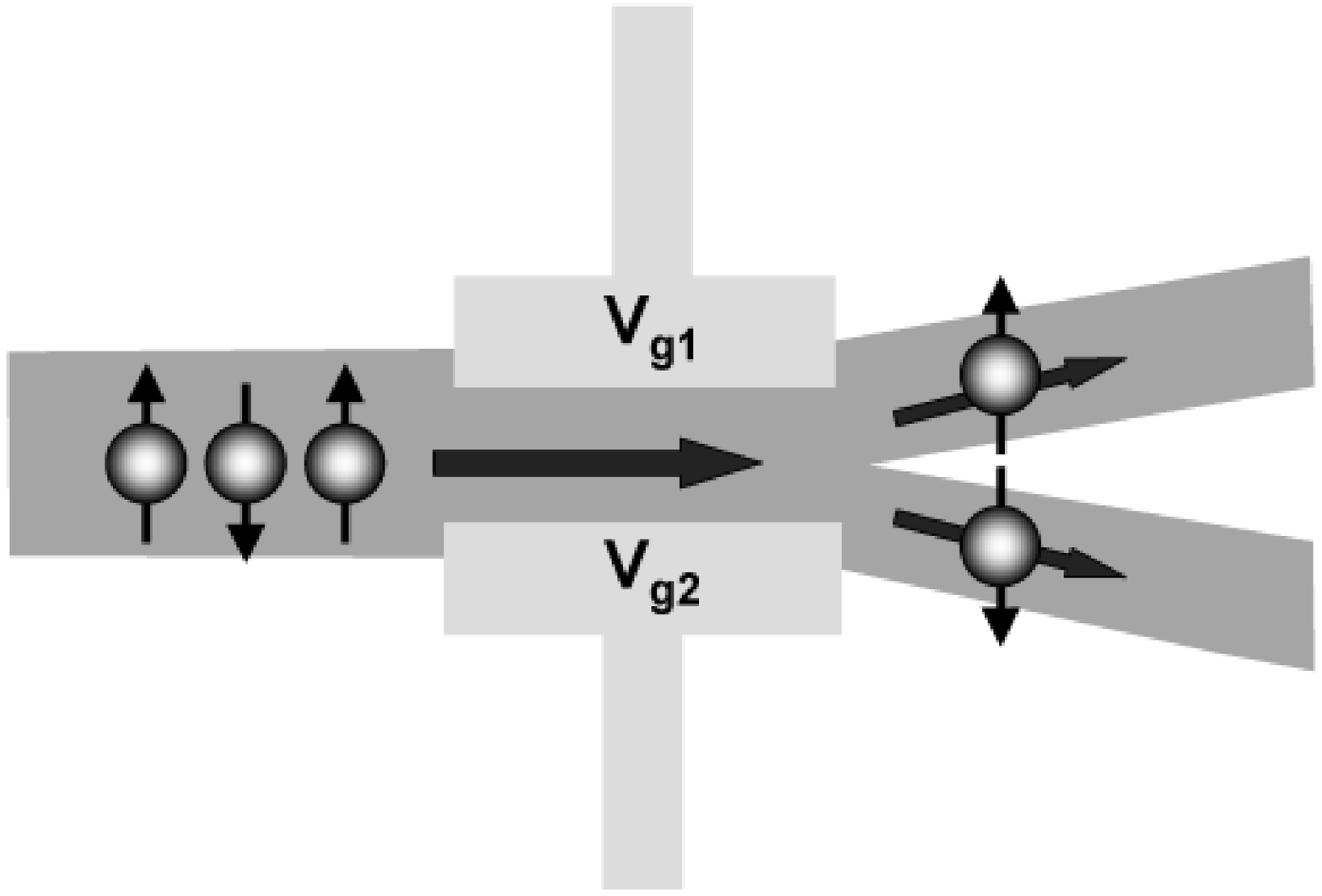}
%\vspace{3cm}
\caption{\label{fig:system}
Top view of Stern-Gerlach spin filter. $V_{g1}$ and $V_{g2}$ are gate voltages
to produce a spatial gradient of spin-orbit interaction. Stern-Gerlach type spin separation 
occurs when un-polarized electrons go through the nonuniform SOI region between the two gate
electrodes. }
\end{center}
\end{figure}

The importance of the modulated SOI on the transport properties
has recently been stressed.\cite{Brouwer,RodriguezA,RodriguezB,Khodas,Sun}
Since the effect of SOI on the propagating electrons is,
in some respect, similar to the effective magnetic field,
the modulated SOI is expected to cause the Stern-Gerlach like
spin separation.
In this paper, we theoretically demonstrate that a mesoscopic
Stern-Gerlach spin filter 
is feasible by using a nonuniform spin-orbit interaction as shown in
Fig.~\ref{fig:system}.
The strength of the SOI is modulated along the direction
perpendicular to the charge current.
We demonstrate that nearly 100\% spin polarization
(perpendicular both to the current and the direction normal to 2DES)
can be achieved without
a ferromagnetic contact and an external magnetic field.
The above results are obtained both from the wave packet dynamics
and from the transfer matrix calculation  of the transmission coefficients.
We have found that the large polarization
is obtained when the electron propagates via the lowest channel
where the transverse mode of the wave function contains only single antinode.
We also investigate the effect of randomness.
The results show
that the polarization of the current survives
as long as the randomness is not so strong.

We consider a 2DES in $x$-$y$ plane and the
current is assumed to flow in the $x$-direction,
while fixed boundary condition is imposed in the $y$-direction.
A square lattice is considered for modeling 2DES 
and only the nearest neighbor hopping
is taken into account.
The tight binding lattice spacing $a$ and the hopping energy
$V_0=\frac{\hbar^{2}}{2m^* a^{2}}$ where $m^*$
is the effective electron mass
are taken as 
the unit length and the unit energy, respectively.
The region
in which the SOI exists is $L_x\times L_y=60 \times 30$,
where $L_x$ and $L_y$ are the length and the width of the system.
We attach ideal leads to both sides of this region.
To include the SOI,
we use the Ando model \cite{Ando} described by the Hamiltonian,
\begin{eqnarray}
H=\sum_{i,\sigma} W_i c^{\dagger}_{i\sigma}c_{i\sigma}
  - \sum_{<i,j>,\sigma,\sigma'} V_{i\sigma,j\sigma'}
    c^{\dagger}_{i\sigma}c_{j\sigma'}
\end{eqnarray}
with
\begin{eqnarray}
V_{i,i+\hat{\bf x}}=
\left(
\begin{array}{cc}
\cos\theta(y) & \sin\theta(y) \\
-\sin\theta(y) & \cos\theta(y)
\end{array}
\right),
\end{eqnarray}
and
\begin{eqnarray}
V_{i,i+\hat{{\bf y}}}=
\left(
\begin{array}{cc}
\cos\theta(y) & -{\mathrm i}\sin\theta(y) \\
-{\mathrm i}\sin\theta(y) & \cos\theta(y)
\end{array}
\right),
\end{eqnarray}
where $c^{\dagger}_{i\sigma}$ $(c_{i\sigma})$ is a creation (annihilation)
operator of an electron on the site $i$ with spin $\sigma$ and
$\hat{{\bf x}} (\hat{{\bf y}})$ the unit vector along $x$-($y$-)direction.
$W_i$ is the random potential distributed uniformly in the
range $[-W/2,W/2]$.  Unless explicitly stated, we consider the
impurity free case.
The strength of SOI is characterized by $\theta (y)$
\begin{eqnarray}
\theta(y)=\frac{2\theta_{\rm max}}{L_y}y-\theta_{\rm max}\,.
\label{theta_y}
\end{eqnarray}
The Hamiltonian of the Rashba SOI \cite{Rashba} is generally described as
\begin{eqnarray}
H_{\rm R}=\frac{1}{\hbar}\{\alpha(y)p_x\sigma_y
-\frac{(\alpha(y)p_y\sigma_x+p_y\alpha(y)\sigma_x)}{2}\},
\label{Hrso}
\end{eqnarray}
where $\alpha(y)$ is the strength of the SOI.
The relation between $\alpha(y)$ and $\theta (y)$ is given by
\begin{eqnarray}
\alpha(y) \simeq 2\theta (y)  V_{0}a
\hspace{1cm} ( \mbox{ for }  \theta (y) \ll 1).
\end{eqnarray}

The spin separation mechanism we propose in this
paper is as follows.  Equation (\ref{Hrso}) means that 
the effective Zeeman field in the $y$-direction
appears when an electron propagates in the $x$-direction.
If the effective Zeeman field has gradient in the $y$-direction,
up and down spin electrons are accelerated in opposite directions.
Unlike the Stern-Gerlach experiment,
this effect is expected to be easily observed even if the particles
have charge.
Since the spins are expected to align in the $y$-direction,
hereafter we concentrate on the polarization in the $y$-direction.

It is possible to make a spatial gradient of
the SOI in the $y$-direction e.g. by using 
two gate electrodes which partially cover the channel, and
the change in the SOI strength between the two electrodes
$\Delta\alpha $ has been experimentally obtained
to be $0.4-0.8\times10^{-11}$ [eVm].\cite{Nitta,Engels}
Let us assume 
$a=30$ [nm] and $m^{\ast}=0.05m_{0}$ where $m_{0}$ is the free electron mass.
Then the system area is $1.8 \times 0.9$ [$\mu {\rm m}^{2}$],
$V_{0} \simeq 0.85$ [meV] and
$\Delta\alpha \simeq 0.64\times10^{-11}$ [eVm]
by setting $\Delta\theta (y) = 0.04\pi$. %2\theta_{\rm max}
We therefore choose $\theta_{\rm max}$ to be $0.02\pi$.

\begin{figure}[htb]
\begin{center}
\includegraphics[scale=0.4,angle=270]{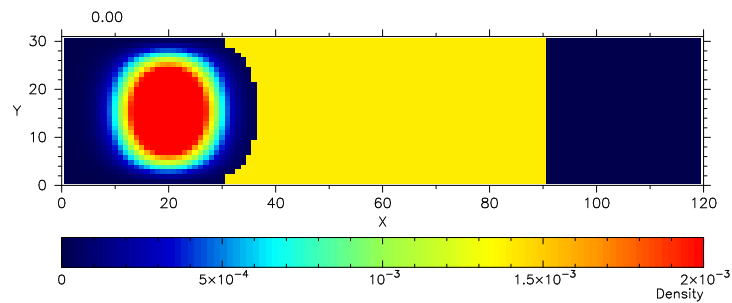}
%\vspace{3cm}
\caption{\label{fig:eqm_ini}
Initial wave packet ($t=0$) 
propagating to the right.
Yellow region indicates the area where SOI is present.
}
\end{center}
\end{figure}

To investigate the electron transport in the nonuniform SOI system,
we calculate the time evolution of the wave packet by the
equation-of-motion method based on the exponential product
formula.\cite{Kawarabayashi}
The charge density 
$\sum_{\sigma} (|\bracket{\uparrow}{\psi_{\sigma}}|^2 +|\bracket{\downarrow}{\psi_{\sigma}}|^2)$
of the initial wave packet is shown in Fig.~{\ref{fig:eqm_ini}}.
The initial wave packet with spin $\sigma$ is assumed to be
\begin{equation}
\psi_{\sigma}(t=0)=A\sin\left(\frac{\pi y}{L_y+1}\right)\, 
\exp\left(
\mathrm{i} k_x x -\frac{\delta k_x^2 x^2}{4}
\right) \chi_\sigma\,,
\label{eq:initialwavepacket}
\end{equation}
with
\begin{equation}
\chi_\uparrow=\frac{1}{ \sqrt{2} }\left(\begin{array}{c}1\\\mathrm{i}\end{array}\right)\,,\,
\chi_\downarrow=\frac{1}{ \sqrt{2} }\left(\begin{array}{c}1\\-\mathrm{i}\end{array}\right)\,\,.
\end{equation}
We set $k_x=0.5$ and $\delta k_x=0.2$.

\begin{figure}[htb]
\begin{center}
\includegraphics[scale=0.4,angle=270]{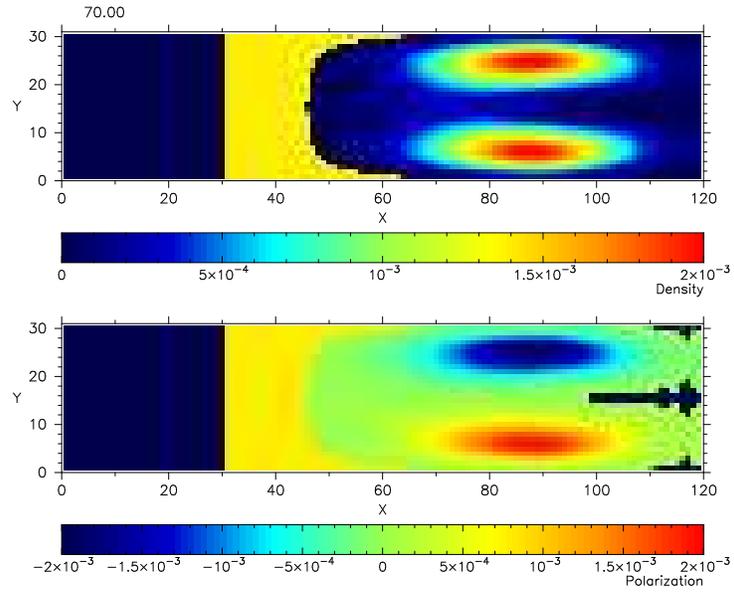}
%\vspace{3cm}
\caption{\label{fig:eqm_fin}
Wave packet after time evolution ($t=70 \cdot \hbar V_{0}^{-1}$).
The strength of the SOI is modulated in the $y$-direction.
Upper: the charge density
$\sum_{\sigma} (|\bracket{\uparrow}{\psi_{\sigma}}|^2 +|\bracket{\downarrow}{\psi_{\sigma}}|^2)$,
and lower: the corresponding polarization,
$\sum_{\sigma} (|\bracket{\uparrow}{\psi_{\sigma}}|^2 -|\bracket{\downarrow}{\psi_{\sigma}}|^2)$.
}
\end{center}
\end{figure}

The wave packet after time evolution is shown in Fig.~\ref{fig:eqm_fin}.
The charge density splits into the upper and the lower parts
with its spin polarizations opposite.
The maximum value of 
$\sum_{\sigma} (|\bracket{\uparrow}{\psi_{\sigma}}|^2 +
|\bracket{\downarrow}{\psi_{\sigma}}|^2)$
%$|\psi_{\uparrow}|^2+|\psi_{\downarrow}|^2$
and that of 
$|\sum_{\sigma} (|\bracket{\uparrow}{\psi_{\sigma}}|^2 -
|\bracket{\downarrow}{\psi_{\sigma}}|^2)|$
%$||\psi_{\uparrow}|^2-|\psi_{\downarrow}|^2|$
are almost the same,
which suggests that nearly 100 $\%$ spin filtering has been achieved.

\begin{figure}[htb]
\begin{center}
\includegraphics[scale=0.75]{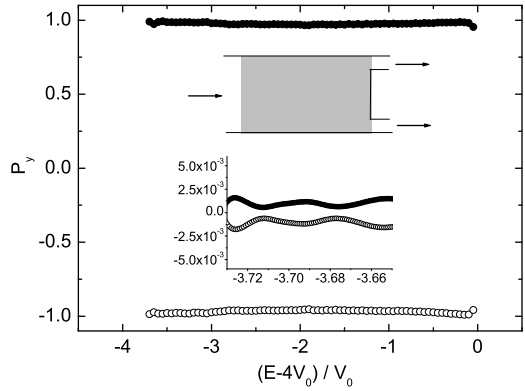}
%\vspace{3cm}
\caption{\label{fig:tmm_ch1}
Polarization $P_y$ (Eq.~(\ref{eq:polarization}))
of the current as a function of the Fermi energy.
Filled (Open) circle shows the polarization 
of the upper (lower) lead in the right hand side.
Main figure shows the component from the lowest channel of the left lead.
Inset shows the result in which the whole channel is taken into account.
The schematic view of the three-terminal geometry is also shown in the inset.
}
\end{center}
\end{figure}

To reinforce the above finding,
we have also calculated the
polarization of the current by the Landauer formula.\cite{Landauer}
Figure~\ref{fig:tmm_ch1} shows the polarization of the
current as a function of the electron Fermi energy $E$
in units of the hopping energy.
Inset of this figure is the schematic view of the system.
Three-terminal geometry has been employed to distinguish
the current through the upper and lower part of the system.
The system area including the nonuniform SOI 
is again $60 \times 30$ and
the width of the upper and lower leads in the right hand side is $5$.
We assume that the chemical potential of the reservoir
attached to the left hand side lead is $E+eV$ with $V$ the voltage,
while both of the upper and lower leads in the right hand side
are attached to the reservoirs with chemical potential $E$.
The transmission coefficients are calculated via
the transfer matrix method \cite{Pendry} extended
to include spins.\cite{ohe03}
In order to realize the multi-terminal system,
the static potential is assumed in the middle of the right leads.
We confirmed that the result of the transfer matrix method
agrees with the result of the Green function method \cite{Dattatxt} up
to 4 digits.

The polarization of the current is defined as
\begin{equation}
P_y=\frac{T_{\uparrow}-T_{\downarrow}}{T_{\uparrow}+T_{\downarrow}},
\label{eq:polarization}
\end{equation}
where $T_{\sigma}= \frac{e^2}{h}
\sum_{\sigma'}|t_{\sigma,\sigma'}|^2$.
Here $t_{\sigma,\sigma'}$ is the transmission coefficient from
the left lead with spin $\sigma'$
to the right leads with spin $\sigma$.
We focus on the component from the lowest channel of the left lead
which corresponds to the result of the time evolution of the wave packet.
As shown in Fig.~\ref{fig:tmm_ch1}
spin filtering effect is clearly obtained in the nonuniform SOI system.
It should be noted that the polarization along $z$-direction can be obtained in
the uniform SOI system by considering multi-terminal geometry.\cite{Kiselev}
However, the polarization has the strong energy dependence and 
the energy region where the large polarization occurs is
very narrow.

In this calculation, we consider the current
in the lowest channel
of the left lead where there are no nodes in the wave function along the
transverse direction  (see Eq.~(\ref{eq:initialwavepacket})). 
In order to obtain the high polarization of the current,
we found that the lowest channel is expected only in the left lead
while the channel mixing is allowed in the sample.
Inset of Fig.~\ref{fig:tmm_ch1} shows the polarization
which takes into account the whole channel.
The spin separation effect of the higher channel
becomes weaker due to the fact that
the transverse wave function has several antinodes along the $y$-direction.
Each antinode splits but the trajectory
is quite different, which
causes the cancellation of the
polarization.
One can avoid this difficulty by attaching
a narrow lead or fabricate a point contact
in the left hand side so that
only the first channel opens.
The complex trajectory is mainly due to the scattering by the
hard wall located to separate the channel to the upper and lower leads.
Using the soft wall potential may improve the situation.

\begin{figure}[htb]
\begin{center}
\includegraphics[scale=0.75]{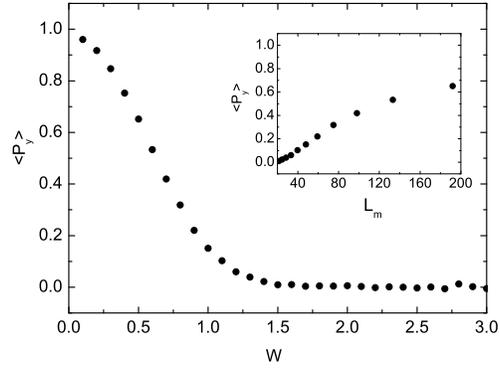}
%\vspace{3cm}
\caption{\label{fig:tmm_ch1_w1}
The polarization of the current of the lower lead
as a function of the strength of disorder.
Inset shows the polarization as a function of the mean free path.
Average over 10000 samples has been performed.}
\end{center}
\end{figure}

From the experimental point of view,
we have to consider the effect of randomness.
Figure~\ref{fig:tmm_ch1_w1} shows that the average of the polarization
in the presence of impurities.
The energy of the electron is fixed to $E/V_0=3.0$ and
average over 10000 samples has been performed.
From this figure, we see that the finite polarization of the current
persists even in the presence of impurities.
Further increase of the randomness significantly
destroys the polarization. 
The strength of the disorder is related to the mean free path in 2DES
without spin-flip scattering as
\begin{equation}
W = \left(
\frac{6\lambda_{F}^{3}}{\pi^{3} a^{2} L_\mathrm{m}}
\right)^{1/2} \cdot E
\end{equation}
where $\lambda_{F}$ denotes the Fermi wavelength and
$L_\mathrm{m}$ the mean free path.\cite{Ando91}
Inset of Fig.~\ref{fig:tmm_ch1_w1} shows the replot of the polarization
as a function of $L_\mathrm{m}$, which indicates that
finite polarization remains as long as $L_\mathrm{m}$
exceeds the system length.
The electron mean free path in 
the InGaAs 2DES channel exceeds $1\mu $m  in the wide
range of the gate voltage,\cite{Engels}
and the spin filtering effect
by nonuniform SOI strength is expected to survive
in actually fabricated devices.
We have also changed the distribution of the impurities to the binary
distribution, where the site energy $W_i$ takes only two values,
0 and $W>0$.  In this case the variance of the potential fluctuation
becomes $p(1-p)W^2$ with $p$ the probability that $W_i$ takes $W$.
We set $p\approx 0.092$ so that $p(1-p)=1/12$ corresponding to
the uniform distribution considered above.
No significant difference is observed by this change of distribution.
We have also confirmed that distributing
attractive impurities, i.e., $W<0$ does not influence the results either.

In conclusion, we have investigated the novel spin filtering
due to the nonuniform spin-orbit interaction (SOI) system.
The strength of SOI is modulated perpendicular to the
charge current, which
yields the gradient of the effective Zeeman field,
and the spin separation occurs as in the Stern-Gerlach experiment.
Both the time evolution of the wave packet and
the transmission coefficient
indicate the large polarization which
survives even in the presence of impurities.
In this mechanism, the direction of the spin polarization can be 
easily switched by the gate voltages between the two gate electrodes. 
When the fully polarized current (which can be produced by the device
proposed here) is injected to this system, one expects that the current
flows along one side of the system and the currents in the upper and lower
leads become totally different, i.e., the information of the spin
polarization is transformed into conductance. Therefore this system can
also be used as a detector of the polarized current.
It should be emphasized that this effect is expected to survive 
even at relatively high temperature.
This is because the mechanism proposed here does not rely on the quantum
interference which is easily destroyed by dephasing.
Also the weak dependence of $P_y$ on energy in the lowest channel
suggests that smearing of the Fermi distribution function is not
important.

The authors are grateful to B. Kramer, K. Dittmer and K. Slevin
for valuable discussions.
This work was supported in part by the Grant-in-Aid from the Ministry of
Education, Culture, Sports, Science and Technology, No. 14540363.
One of the authors (J.O.) was supported by Research Fellowships of
the Japan Society for the Promotion of Science for Young Scientists.

%\vspace{2cm}

\end{document}